\documentclass[aps, twocolumn, 10pt, showpacs, longbibliography, superscriptaddress]{revtex4-1}
\usepackage{fullpage}
\usepackage{graphicx}
\usepackage{dcolumn}
\usepackage{bm}
\usepackage{color}
\usepackage{amsmath, amsthm, amssymb}
\usepackage[modulo]{lineno}
\usepackage[usenames,dvipsnames]{xcolor}
\usepackage{ulem}
\usepackage{epstopdf}
\newcommand{\ba}{\begin{eqnarray}}
\newcommand{\ea}{\end{eqnarray}}

\usepackage{float}
\floatstyle{boxed}

\begin{document}
\title{Graphene-based Josephson junction microwave bolometer}
\author{Gil-Ho Lee}
\affiliation{Department of Physics, Harvard University, Cambridge, MA 02138}
\affiliation{Department of Physics, Pohang University of Science and Technology, Pohang 790-784, Republic of Korea}
\author{Dmitri K. Efetov}
\affiliation{ICFO-Institut de Ci\`{e}ncies Fot\`{o}niques, The Barcelona Institute of Science and Technology, 08860 Castelldefels (Barcelona), Spain}
\author{Woochan Jung}
\affiliation{Department of Physics, Pohang University of Science and Technology, Pohang 790-784, Republic of Korea}
\author{Leonardo Ranzani}
\affiliation{Raytheon BBN Technologies, Quantum Information Processing Group, Cambridge, Massachusetts 02138, USA}
\author{Evan D. Walsh}
\affiliation{Department of Electrical Engineering and Computer Science, Massachusetts Institute of Technology, Cambridge, MA 02139}
\affiliation{School of Engineering and Applied Sciences, Harvard University, Cambridge, MA 02138}
\author{Thomas A. Ohki}
\affiliation{Raytheon BBN Technologies, Quantum Information Processing Group, Cambridge, Massachusetts 02138, USA}
\author{Takashi Taniguchi}
\affiliation{National Institute for Materials Science, Namiki 1-1, Tsukuba, Ibaraki 305-0044, Japan}
\author{Kenji Watanabe}
\affiliation{National Institute for Materials Science, Namiki 1-1, Tsukuba, Ibaraki 305-0044, Japan}
\author{Philip Kim}
\affiliation{Department of Physics, Harvard University, Cambridge, MA 02138}
\author{Dirk Englund}
\affiliation{Department of Electrical Engineering and Computer Science, Massachusetts Institute of Technology, Cambridge, MA 02139}
\author{Kin Chung Fong}
\email{fongkc@gmail.com}
\affiliation{Raytheon BBN Technologies, Quantum Information Processing Group, Cambridge, Massachusetts 02138, USA}
\date{\today}
\maketitle

\textbf{
Sensitive microwave detectors are critical instruments in radioastronomy \cite{Benford:2004tb}, dark matter axion searches \cite{Graham:2015ib}, and superconducting quantum information science \cite{Govia:2014jq, Inomata:2016jc}. The conventional strategy towards higher-sensitivity bolometry is to nanofabricate an ever-smaller device to augment the thermal response \cite{Wei:2008jw, Gasparinetti:2015gr, Govenius:2016fo}. However, this direction is increasingly more difficult to obtain efficient photon coupling and maintain the material properties in a device with a large surface-to-volume ratio. Here we advance this concept to an ultimately thin bolometric sensor based on monolayer graphene. To utilize its minute electronic specific heat and thermal conductivity, we develop a superconductor-graphene-superconductor (SGS) Josephson junction  \cite{Lee:2011et, Coskun:2012hx, Borzenets:2013vz, Calado:2015fp, BenShalom:2015hy, Wang:2019dy} bolometer embedded in a microwave resonator of resonant frequency 7.9 GHz with over 99\% coupling efficiency. From the dependence of the Josephson switching current on the operating temperature, charge density, input power, and frequency, we demonstrate a noise equivalent power (NEP) of 7~$\times 10^{-19}$~W/Hz$^{1/2}$, corresponding to an energy resolution of one single photon at 32 GHz \cite{Moseley:1984wh} and reaching the fundamental limit imposed by intrinsic thermal fluctuation at 0.19 K.
}

\begin{figure*}  
\includegraphics[width=1.8\columnwidth]{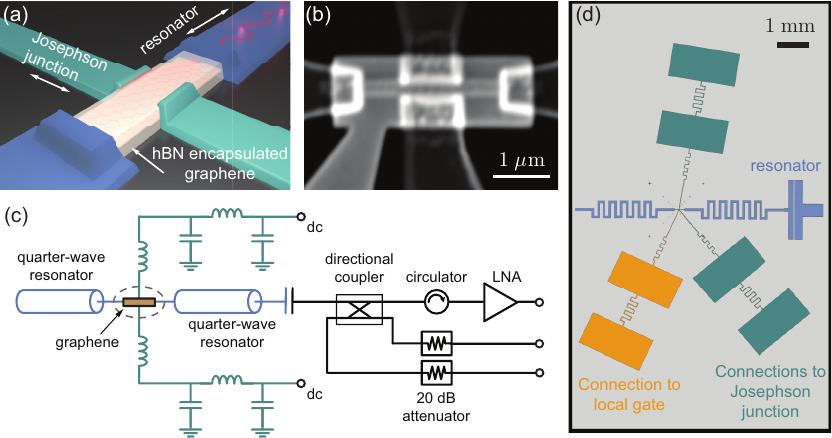}
\caption{(a) Device concept of the superconducting-graphene-superconducting (SGS) Josephson junction (JJ) microwave bolometer. The hBN-encapsulated SGS JJ (1 $\mu$m wide and with a gap of $\simeq$0.3 $\mu$m) is embedded simultaneously in a half-wave resonator to allow microwave coupling (blue) and DC readout (green) of the JJ. For clarity, the local gate is not shown. (b) Scanning electron microscope image of the orthogonal-terminal JJ. (c) Schematics of the detector setup. The graphene flake is located at the current antinode of the half-wave microwave resonator. Test microwave power is coupled to the detector through the 20 dB directional coupler and highly attenuated coaxial cables from room temperature. Two stages of inductors and capacitors form a low-pass filter network for the DC measurement. (d) False-colored optical image of the actual device.}
\end{figure*}

Many attractive electrical and thermal properties in graphene make it a promising material for bolometry and calorimetry \cite{Vora:2012cs, Fong:2012ut, Yan:2012eh, McKitterick:2013ue, Efetov:2018ij, Han:2013cg, Cai:2014vn, ElFatimy:2016fp}. It can absorb photons from a wide frequency bandwidth efficiently by impedance matching \cite{Walsh:2017kk}; the electron-electron scattering time is short and can quickly equilibrate the internal energy from absorbed photons to evade leakage through optical phonon emission \cite{Tielrooij:2013cd}; its weak electron-phonon coupling can keep the electrons thermally isolated from the lattice \cite{Viljas:2010jj, Song:2012bya, Chen:2012et, Betz:2013up, Vora:2012cs, Fong:2012ut, Borzenets:2013vz, Graham:2013vl}; most importantly, at the charge neutrality point (CNP), graphene has a vanishing density of states. This results in a small heat capacity and electron-to-phonon thermal conductance which are highly desirable material properties for bolometers and calorimeters, while maintaining a short thermal response time \cite{Efetov:2018ij}. Although the bolometric response of graphene has been tested in devices based on noise thermometry \cite{Fong:2012ut, McKitterick:2013ue, Efetov:2018ij}, their performance is severely hampered by the degrading thermometer sensitivity when the electron temperature rises upon photon absorption \cite{McKitterick:2013ue}. Here, we overcome this challenge by adopting a fundamentally different measurement technique: we integrate monolayer graphene simultaneously into a microwave resonator and a Josephson junction, and upon absorbing microwave radiation into the resonator, the rise of the electron temperature in graphene suppresses the switching current of the SGS Josephson junction. This mechanism can function as the bolometer readout and provide us a way to study the thermal response of this bolometer.

Inspired by the demonstration of using heating or quasiparticle injection to control the supercurrent in superconductor-normal-superconductor junctions in the DC regime \cite{Tirelli:2008km, Morpurgo:1998bp}, we design our microwave bolometer with a orthogonal-terminal graphene-based Josephson junction (GJJ) as shown in Fig. 1a and b. The monolayer graphene is encapsulated on the top and bottom by hexagonal boron-nitride (hBN). The proximitized Josephson junction (green color) is formed by edge-contacting NbN superconductors to the graphene such that dissipationless Josephson current can flow along the JJ direction \cite{Calado:2015fp}. A dissipative microwave current can flow along the direction perpendicular to that of the junction, with the graphene extended out by 0.8 $\mu$m from each side of the GJJ before connecting to quarter-wave resonators (blue color) to form a half-wave resonator using a NbN microstrip with a characteristic impedance of  86 $\Omega$ (Fig. 1c and d). This extension is narrow and long to prevent Josephson coupling to the microstrips and positions the graphene at the current antinode of the resonator.

Microwave power is applied to the resonator through a 200 fF coupling capacitor. We can characterize our GJJ-embedded resonator by reflectometry using a directional coupler. All test power is delivered via the heavily-attenuated microwave coaxial cables to filter the thermal noise from room temperature. To decouple the GJJ DC measurement from the microwave resonator, two stages of LC low-pass filters are implemented to form a high-impedance line at high frequency. The 1 nH inductors are made of narrow meandered wires and are shunted by 530 fF capacitor plates. 

\begin{figure}[t!]  
\includegraphics[width=\columnwidth]{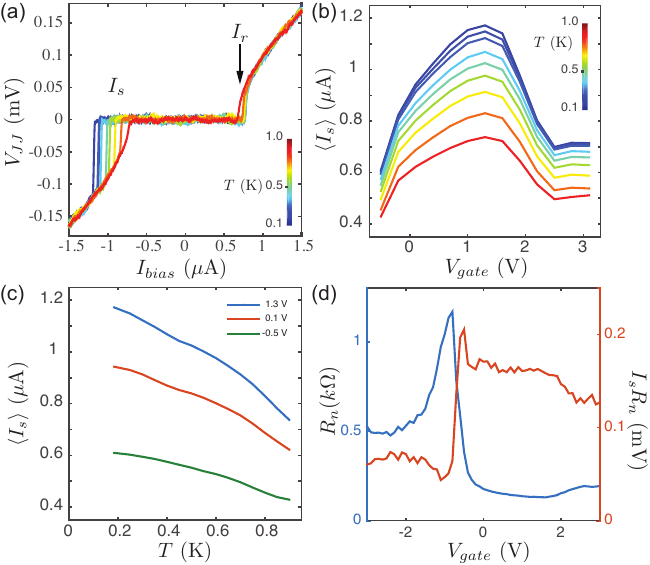}
\caption{Characterizing the graphene-based Josephson junction (GJJ) switching current. (a) GJJ voltages with sweeping of bias current and (b) the averaged switching current vs. gate voltages at various temperatures, 0.19-0.9 K. (c) Averaged switching currents vs. temperatures at different gate voltages. (d) GJJ normal resistance and the $\langle I_{\rm{s}}\rangle R_{\rm{n}}$ product as a function of gate voltage.}
\end{figure}

We study the GJJ switching as a function of temperature and gate voltage. Fig. 2a shows the typical voltage drop across the junction $V_{\rm{JJ}}$ as the DC bias current is swept from 1.5 to -1.5 $\mu$A at device temperatures between 0.19 and 0.9 K. Our GJJ shows hysteretic switching behavior: the switching current $I_{\rm{s}}$, at which the junction switches from the dissipationless state to the normal state, is different from the retrapping current, $I_{\rm{r}}$. Such hysteresis is presumably due to self-Joule heating when the junction turns normal \cite{Borzenets:2013vz}. The averaged switching currents $\langle I_{\rm{s}}\rangle$ are plotted at various gate voltages $V_{\rm{gate}}$ and temperatures in Figs. 2b and c. The drop of $\langle I_{\rm{s}}\rangle$ as temperature rises is an important feature that can determine the sensitivity of the GJJ as a bolometer as well as the quantum efficiency and dark count of the future microwave single photon detector \cite{Walsh:2017kk}. Fig. 2d plots the normal-state junction resistance $R_{\rm{n}}$ as a function of gate voltage, indicating that the CNP is at -0.9 V. We note that the unusual rise of $R_{\rm{n}}$ at around 2 to 3 V of $V_{\rm{gate}}$ may be due to the formation of a Moir\'{e} superlattice with the hBN substrate (see Method). The $\langle I_{\rm{s}}\rangle R_{\rm{n}}$ product is on the order of 0.16 mV, which is comparable to other GJJs of similar size in the long diffusive limit \cite{Walsh:2017kk}.

\begin{figure}  
\includegraphics[width=0.8\columnwidth]{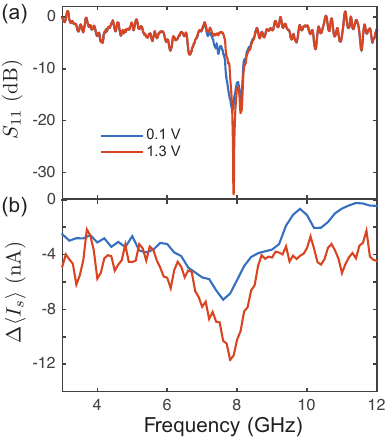}
\caption{Demonstration of the device's operation as a bolometer and measuring the detector efficiency. (a) The scattering parameter of the device at 0.19 K shows a resonance near 7.9 GHz with linewidths of 861 and 599 MHz at gate voltages of 0.1 and 1.3 V respectively, obtained from the quality factor analysis. (b) Suppression of the average switching current at -112 dBm relative to the absence of input power at two different gate voltages.}
\end{figure}

The coupling efficiency can be characterized using reflectometry (see Fig. 3a). We design the resonator to be critically coupled at about 7.9 GHz. The dissipation is dominated by the monolayer graphene, which can be modeled as a resistor located at the current antinode. At -112 dBm input power, we measured the change in $\langle I_{\rm{s}}\rangle$ as a function of input frequency for two different gate voltages, 0.1 and 1.3 V, with $\langle I_{\rm{s}}\rangle$ of 0.94 and 1.17 $\mu$A respectively (see Fig. 3b). If the absorbed microwave photon caused a resonant excitation at the plasma frequency of the GJJ, we would expect the frequency of the resonance dip in $\langle I_{\rm{s}}\rangle$ to shift by approximately 1 GHz, because the GJJ plasma frequency is proportional to the square root of the critical current. In contrast, the suppression of $\langle I_{\rm{s}}\rangle$ aligns closely to the microwave resonant frequency measured by reflectometry. Therefore we cannot attribute the suppression of $\langle I_s\rangle$ to resonant excitation of the GJJ. We note that the linewidths in Fig. 3b match to the one given by the loaded quality factors of 9 and 13 at $V_{\rm{gate}}$ = 0.1 and 1.3 V respectively, obtained from the fitting of the phase of the scattering parameter in Fig. 3a.

\begin{figure}  
\includegraphics[width=\columnwidth]{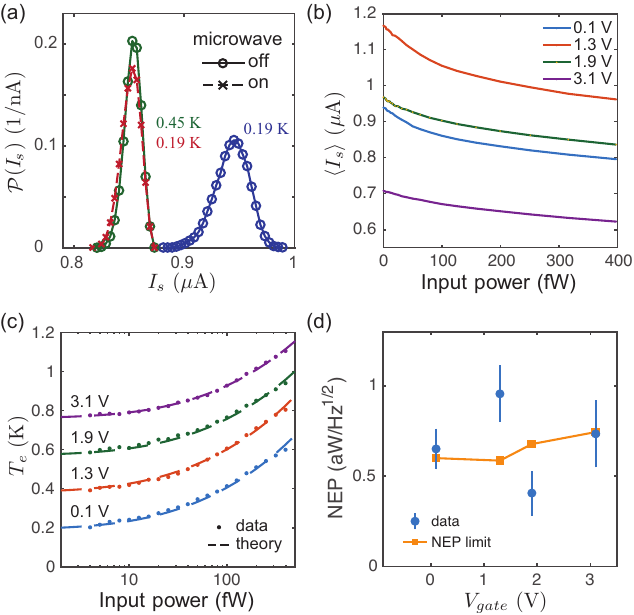}
\caption{Sensitivity and the fundamental fluctuation limit of the bolometer. (a) Distributions of the switching current at 0.19 K (blue circles) and 0.45 K (green circles) without microwave input, and at 0.19 K with microwave input power of 126 fW (red crosses) at $V_{\rm{gate}}$ = 0.1 V. (b) Average switching current as a function of microwave input power at 7.9 GHz at various gate voltages. (c) Interpolated graphene electron temperature using the results in Fig. 2c with 0.19 K offsets for clarity. The dashed lines are fits to the theory of heat transfer from electrons to phonons. (d) Measured noise equivalent power (NEP) and the fluctuation limit. The averaged NEP is 0.7 $\pm$ 0.2 aW/Hz$^{1/2}$, corresponding to an energy resolution of a single 32 GHz photon.}
\end{figure}

Supercurrent switching statistics can reveal the basic properties of the GJJ \cite{ Lee:2011et, Coskun:2012hx}, and hence its thermal response as a bolometer. We measure the distribution of $I_{\rm{s}}$ by recording the potential drop across the GJJ while sweeping the bias current for 6,000 times per gate voltage, input power, and temperature. The decay of the GJJ from the supercurrent state to the normal state is stochastic and the typical distribution is plotted in Fig. 4a at two temperatures, 0.19 and 0.45 K, at $V_{\rm{gate}}$ of 0.1 V without power input. The decay rate (also known as the escape rate from the tilted-washboard potential in the resistively and capacitively shunted junction model), and thus the switching probability, can be determined uniquely from the distribution using the Fulton-Dunkleberger method \cite{Lee:2011et, Coskun:2012hx}. When the experiment is conducted at 0.19 K with an increasing power input, the switching histogram shifts gradually to lower values. When the microwave input power reaches 126 fW when the device is at 0.19 K, the distribution overlaps well to that at 0.45 K with zero input power and therefore the GJJ has the same decay rate under these two conditions. This suggests the suppression of $\langle I_{\rm{s}}\rangle$ is due to the heating of graphene electrons from 0.19 K to 0.45 K by the microwave input power, instead of other mechanisms such as the AC Josephson effect or an additional bias current across the GJJ \cite{Oelsner:2017er}.

$\langle I_{\rm{s}}\rangle$ decreases monotonically as we raise the microwave input power $P_{\rm{mw}}$. Fig. 4b plots $\langle I_{\rm{s}}\rangle$ as a function of input power. Using the measured $\langle I_{\rm{s}}\rangle$ at various device temperatures in Fig. 2b, we can use an interpolation to calculate graphene electron temperature $T_e$ as a function of $\langle I_s\rangle$ which is a function of $P_{\rm{mw}}$. The results $T_e(P_{\rm{mw}})$ are shown in Fig. 4c for four different gate voltages with offsets in multiples of 0.19 K in the y-axis for clarity. The dashed lines plot the fits to the data using the electron-phonon heat transfer equation \cite{Viljas:2010jj, Song:2012bya, Chen:2012et}: $P = \Sigma A\left( T_e^\delta - T^\delta\right)$ where $A$ is the area of the monolayer graphene, and $\Sigma$ and $\delta$ are the electron-phonon coupling parameter and its temperature power law, respectively. The best fitted $\Sigma$ are 2.14, 2.04, 2.74, and 3.30 Wm$^{-2}$K$^{-3}$ in the ascending order of gate voltages with $\delta = 3$. This temperature power law corresponds to the cooling of graphene electrons mediated by supercollision or disorder \cite{Song:2012bya, Chen:2012et, Betz:2013up, Graham:2013vl}. However, using the deformation potential of 20 eV and measured electron mobility of 20000 cm$^2$/Vs from the GJJ normal resistance versus the gate voltage, the same theory predicts a $\Sigma$ of 0.0086 Wm$^{-2}$K$^{-3}$. This large discrepancy suggests that the existing electron cooling theories of the defect-mediated electron-phonon coupling are not applicable when $l_{\rm{mfp}}$ (340 nm in our sample) is larger or comparable to the sample dimensions (0.3 $\mu$m by 2.6 $\mu$m in our device). A recent scanning nanothermometry experiment \cite{Halbertal:2017je} has spatially imaged the cooling of electrons in high-quality graphene and demonstrated that the cooling of electrons can be dominated by the atomic defects on the edge rather than those in the bulk. Therefore, we can expect $l_{\rm{mfp}}$ based on the bulk electrical transport may underestimate the total cooling rate of the electron-phonon coupling when the $l_{\rm{mfp}}$ is larger than the sample size. Since these scatterings by atomic defects on edge scales with the sample perimeter whereas the defect-mediated scattering scales with sample area, more systematic experiments with consistently etched graphene flakes of different sample aspect ratios can provide more understanding of the cooling of electrons to achieve a higher-sensitivity graphene-based bolometer in the future.

The effectiveness of the thermal insulation at the graphene-superconductor contacts due to Andreev reflection can be evaluated using the Wiedemann-Franz law. If there is any heat diffusion at the contacts, the GJJ (being wide and short) will have the largest contribution to the thermal conductance which, based on the one-dimensional thermal model \cite{Fong:2012ut}, is given by $4(\pi k_{\rm{B}}/e)^2(T/R)$ where $k_{\rm{B}}$ is the Boltzmann constant, $e$ is the electron charge, and $R$ is the electrical resistance between the contacts. For $R_{\rm{n}}$ = 145 $\Omega$ at $V_{\rm{gate}}$ = 1.9 V, this would be about 387 pW/K and is 1000 times larger than the measured thermal conductance $G_{\rm{th}} = \delta\Sigma AT^{\delta-1}$, i.e. 230 fW/K, at 0.19 K. This suggests the NbN superconductor used in the experiment is acting as a good thermal insulator, prohibiting the heat diffusion at the graphene-superconductor interface.

We can estimate the NEP by first identifying the minimum input power $\delta P_{\rm{min}}$ required to suppress the $\langle I_{\rm{s}}\rangle$ by one standard deviation of the switching distribution $\sigma_{\langle I_{\rm{s}}\rangle}$. It is given by $\sigma_{\langle I_{\rm{s}}\rangle}\cdot \left(\left| d\langle I_{\rm{s}}\rangle/dP_{\rm{mw}}\right|_{P_{\rm{mw}}=0}\right)^{-1}$, where $d\langle I_{\rm{s}}\rangle/dP_{\rm{mw}}$ is the slope in Fig. 4b, i.e. $\delta P_{\rm{min}}$ = 11.4 fW for $\sigma_{\langle I_{\rm{s}}\rangle}$ of 13.2 nA at 0.19 K and $V_{\rm{gate}}$ = 1.9 V. Then we need to consider the time duration required to detect this $\delta P_{\rm{min}}$ by comparing three time scales: resonator input coupling rate, resonator dissipation rate, and thermal time constant $\tau_{\rm{th}}$. Analyzing the scattering parameter of the resonator shows the coupling and internal quality factors of 12.8 and 10.0, respectively (see Method). The resonator is nearly critically coupled with coupling rate and dissipation rate $\Delta f_{\rm{int}}$ of 630 and 790 MHz, respectively. The thermal time constant is given by the ratio of the graphene electron heat capacitance $C_e$ to $G_{\rm{ep}}$. Since $C_e = A\gamma T$ \cite{Viljas:2010jj} where $\gamma = (4\pi^{5/2}k_{\rm{B}}^2n^{1/2})/(3hv_{\rm{F}})$ is the Sommerfeld coefficient with $n$ as the electron density, $h$ as Planck's constant, and $v_{\rm{F}}$ as the Fermi velocity of electrons in graphene. At 0.19 K and $n \simeq 2\times 10^{12}$ cm$^{-2}$, $C_e \sim$10 $k_{\rm{B}}$ resulting in $\tau_{\rm{th}} \simeq$ 0.6 ns. Therefore, the fastest GJJ detection time is bounded by the rate at which the resonator dissipates energy into the graphene such that our GJJ bolometer has a NEP of $\delta P_{\rm{min}}/\sqrt{\Delta f_{\rm{int}}}$. The result is plotted in Fig. 4d and the error bar is dominated by the accuracy in obtaining $d\langle I_{\rm{s}}\rangle/dP_{\rm{mw}}$ and $\Delta f_{\rm{int}}$. The NEP achieved by this SGS bolometer is, on average across different $V_{\rm{gate}}$, 0.7 $\pm$ 0.2 aW/Hz$^{1/2}$.

Compared to the state of the art, our GJJ shows promise for a range of applications. The GJJ bolometer can operate $\sim10^5$ times faster than the nanowire for its shorter $\tau_{\rm{th}}$, making GJJ bolometer an attractive component for ultrawide-IF-bandwidth hot-electron-bolometric mixer. It also has a much lower energy resolution, equivalent to a single-32 GHz-photon energy \cite{Moseley:1984wh}, because of the small $C_e$ \cite{Vora:2012cs, Fong:2012ut, McKitterick:2013ue, Efetov:2018ij}. Unlike the superconducting-qubit-based and nanowire SPDs, the GJJ detector does not require qubit state preparation nor does it rely on the breaking of Cooper pairs to generate a detectable signal, making it suitable for continuous photon sensing over a wide photon energy range.

Intrinsic thermal fluctuation of a canonical ensemble imposes a fundamental limit on the sensitivity of a bolometer given by $\sqrt{4G_{\rm{th}}k_{\rm{B}}T^2}$ \cite{Moseley:1984wh}. Comparison of the data in Fig. 4d suggests that the NEP of our bolometer as predicted by such fluctuation (based on the measurement of the electron-phonon coupling) is in close agreement to the NEP that we measure using the suppression of switching current resulting from microwave input power. This also suggests that $1/\tau_{\rm{th}}$ is nearly the same as the internal dissipation rate of the resonator. The same temperature scaling law projects a further improvement to 10$^{-21}$ W/Hz$^{1/2}$ at 20 mK. The same detector design could perform calorimetry to detect single microwave photons with further optimization of $d\langle I_{\rm{s}}\rangle/dT$ \cite{Walsh:2017kk}. For a continuous power readout while keeping the GJJ non-dissipative in the supercurrent state, we can employ an RF resonance readout to detect the change of the Josephson inductance of the GJJ.

\textbf{Method.} The device is fabricated by first encapsulating the monolayer graphene between two layers of atomically flat and insulating boron nitride ($\simeq$30 nm thick) using the dry-transfer technique. The superconducting terminals consist of 5-nm-thick niobium and 60-nm-thick niobium nitride deposited after reactive ion etch and electron beam deposition of 5-nm titanium to form the one-dimensional contact \cite{Calado:2015fp, Walsh:2017kk}. Finally, we make the local gate to control the carrier density of the monolayer graphene by growing an aluminium oxide dielectric layer by atomic layer deposition and depositing a layer of gold electrode, before wiring it through the reactive low-pass filter to provide isolation to the microwave circuit.

We design the device for optimal impedance matching to a 2 k$\Omega$ graphene resistance, an estimated value based on its dimensions. Energy dissipation is dominated by Joule heating into the graphene in such a structure, since the typical internal Q-factor of NbN superconducting resonators without a graphene flake is on the order of $10^5-10^6$, compared to the internal Q-factor of our device measured to be less than 30, based on the circle fitting method \cite{Khalil:2012jr}. We achieve optimal impedance matching at critical coupling, where the resonator internal Q-factor due to the graphene resistance is equal to the coupling Q-factor, by adjusting the coupling gap capacitor. We simulate the device with different gap capacitor values using a Method of Moments electromagnetic simulator and determine a coupling capacitor value of 200~fF.

\textbf{Acknowledgements.} We thank valuable discussions with L. Levitov, M.-H. Nguyen, and W. Kalfus. G.-H.L. acknowledges Samsung Science and Technology Foundation under Project Number SSTF-BA1702-05 and National Research Foundation of Korea (NRF) Grant funded by the Korean Government (No. 2016R1A5A1008184). D.K.E. acknowledges support from the Ministry of Economy and Competitiveness of Spain through the ``Severo Ochoa'' program for Centres of Excellence in R\&D (SE5-0522), Fundaci\'{o} Privada Cellex, Fundaci\'{o} Privada Mir-Puig, the Generalitat de Catalunya through the CERCA program, the H2020 Programme under grant agreement 820378, Project: 2D$\cdot$SIPC and the La Caixa Foundation. The work of E.D.W. and D. E. was supported in part by the Army Research Laboratory Institute for Soldier Nanotechnologies program W911NF-18-2-0048 and the US Army Research Laboratory (Award W911NF-17-1-0435). K.W. and T.T. acknowledge support from the Elemental Strategy Initiative conducted by the MEXT, Japan, A3 Foresight by JSPS and the CREST (JPMJCR15F3), JST. K. C. F. was  supported by the internal research of Raytheon BBN Technologies and in part by Army Research Office under Cooperative Agreement Number W911NF-17-1-0574.

\textbf{Appendix.}
\textit{Contact resistance.} We can consider how the contact resistance may impact the device (1) on the Josephson junction measurement and (2) on the microwave resonator. Nyquist noise by the graphene/NbN contact in the junction direction (vertical direction in Fig. 1b-d) would be absent as the bolometer operates in the supercurrent regime where two-probe resistance is zero. Nyquist noise would come into play only after the JJ switches to the resistive regime by microwave photon absorption and gives finite normal resistance of $R_n$. Therefore, we do not need to consider Nyquist noise for determining the NEP of the bolometer. The noise due to thermal and quantum fluctuation on the current-biased Josephson junction are included in determining the NEP because both of these noise contributes to the width of the switching current distribution in Fig. 4a. They are the mechanisms in the thermal activation and macroscopic quantum tunneling of the phase particle of the current-biased Josephson junction \cite{MDC1987PRB} and can be determined using the Fulton-Dunkleberger method \cite{Fulton:1974uf}.

On the other hand, it is possible that the contact resistance between the graphene flake and microwave resonator degrade the bolometer by dissipating the photon energy at the contact instead of the graphene flake. To estimate this effect, let the graphene resistance along resonator direction be $R_{n,res}$ and the graphene/NbN contact resistance along the resonator (horizontal direction in Fig. 1b-d) be $R_c$. We have measured normal resistance $R_n \simeq 145~\Omega$ along the junction direction that has width $W$ = 1 $\mu$m and length $L$ = 270 nm, so square resistance $R_{sq} = R_n/(L/W)$. With the graphene width $W^\prime$ = 300 nm and length $L^\prime$ = 2.6 $\mu$m along the resonator direction, we can roughly estimate $R_{n,res} =R_{sq}*(L^\prime/W^\prime) \simeq 5$ k$\Omega$. If we assume that the contact transparencies for graphene/NbN interfaces along the junction direction and along the resonator direction are similar since the NbN for both the GJJ and the resonator was deposited at the same time, $R_c$ can be estimated by $R_c=(\pi/k_FW^\prime)*(h/4e^2)/T\simeq 0.4$ k$\Omega$, where $k_F$ is Fermi wavenumber and $T$ = 0.8 is a contact transparency estimated from the relationship of $R_n=(\pi/k_FW)*(h/4e^2)/T$ for ballistic graphene channel along JJ direction (see the next paragraph for the discussion on ballistic nature of graphene in our experiment). Contact contribution to the photon energy dissipation by $R_c$ is less than 10\% of total resistance given by $R_{n,res}+R_c$. Thus, we expect that the contact resistance would not significantly degrade bolometer performance.

The graphene-based Josephson junction is at or nearly at the ballistic limit. This is because if we assume the graphene is in the diffusive regime, the Drude mobility and mean free path of graphene are estimated to be 20,000 cm$^2$/Vs and 340 nm, respectively. However this mean free path exceeds the junction length of 270 nm. This is usually the case when graphene is encapsulated by atomically flat and insulating hBN flakes and protected from dirty environment during the fabrication processes. Ref. \cite{Ponomarenko:2013hl} describes how the formation of a Moir\'{e} superlattice with the hBN substrate can give rise to the unusual rise of $R_{\rm{n}}$ at around 2 to 3 V of $V_{\rm{gate}}$.

\textit{Electron-phonon cooling.} We can use the electron-phonon coupling theory in the supercollision or disorder regime to calculate $\Sigma$ \cite{Song:2012bya, Chen:2012et}: \begin{eqnarray}\Sigma = \frac{2\zeta(3)}{\pi^2}\frac{E_{\rm{F}}}{v_{\rm{F}}^3\rho_{\rm{M}}}\frac{\mathcal{D}^2k_{\rm{B}}^3}{\hbar^4l_{\rm{mfp}}s^2}\end{eqnarray} where $\zeta$ is the Zeta function, $E_{\rm{F}}$ is the Fermi energy of graphene charge carriers, $v_{\rm{F}}$ is the Fermi velocity in graphene, $\rho_{\rm{M}}$ is the mass density of the graphene sheet, $\mathcal{D}$ is the deformation potential, $k_{\rm{B}}$ is the Boltzmann constant, $\hbar$ is the reduced Planck constant, $l_{\rm{mfp}}$ is the mean-free-path, and $s$ is the sound velocity of graphene lattice. However, the enhanced electron-phonon cooling that we observed is more likely due to the resonant-scattering by defects located around the edge of graphene flake \cite{Halbertal:2017je, Kong2018, Draelos2019}. $\Sigma$ values are listed in Table I.

$\Sigma$ is independent of charge carrier density if we assume the electron mobility $\mu_e$ is a constant of carrier density such that, with $e$ as the electron charge, and $\tau$ and $m$ as the scattering time and mass of the charge carriers, respectively, $ \mu_e = e \tau/m = e v_{\rm{F}} l_{\rm{mfp}}/E_{\rm{F}}$. Measured electrical transport and Josephson junction parameters are listed in Table II.

\begin{table*}
\begin{tabular}{ l  c  c  c  c} 
\hline
$V_{gate}$ (V)&~~~~~~~0.1~~~~~~~&~~~~~~~1.3~~~~~~~&~~~~~~~1.9~~~~~~~&~~~~~~~3.1~~~~~~~\\ 
\hline
carrier density ($10^{12}$ cm$^{-2}$) & 0.72 & 1.6 & 2.0 & 2.9\\
$R_n$ ($\Omega$) & 160 & 127 & 145 & 195 \\
$C_e$ ($k_B$) & 6.1 & 9.0 & 10 & 12\\
$\Sigma A$ ($\times 10^{-12}$ WK$^{-1}$) from fitting & 1.67 & 1.59 & 2.13 & 2.57\\
$\Sigma$ (Wm$^{-2}$K$^{-1}$) from fitting & 2.1 & 2.0 & 2.7 & 3.3\\
$\Sigma$ (Wm$^{-2}$K$^{-1}$) from theory &  \multicolumn{4}{c}{0.0086}\\
$G_{th}$ (fW/K) & 181 & 173 & 231 & 279\\
$\langle I_s\rangle$ ($\mu$A) & 0.943 & 1.17 & 0.978 & 0.714\\
$\sigma_{I_s}$ (ns) & 15.0 & 23.7 & 13.2 & 9.96\\
$\left| d\langle I_s\rangle /dP\right|$ ($10^6 A/W$) & 1.1 & 1.5 & 1.2 & 0.43\\
$\delta P_{min}$ (fW) & 13.5 & 15.9 & 11.4 & 23.3\\
$Q_{int}$ & 18.3 & 28.5 & 10.0 & 7.9\\
$Q_{couple}$ & 18.4 & 24.4 & 12.8 & 9.7\\
resonator internal dissipation rate (MHz) & 432 & 277 & 790 & 1000\\
NEP ($\times 10^{-19}$ W/Hz$^{-1/2}$)& 6.5 & 9.6 & 4.1 & 7.4\\
\hline
\end{tabular}
\label{NEP}
\caption{List of parameters to estimate NEP and thermal properties of GJJ bolometer in this report. Data refers to operation at 0.19 K with graphene area of 0.78 $\mu$m$^{-2}$ and $V_{CNP} = -0.9$ V unless stated otherwise.}
\end{table*}

\begin{table}
\centering
\begin{tabular}{ l c } 
\hline
 \multicolumn{2}{c}{\textbf{Josephson junction parameters at $V_{gate} = 1.9$ V }}\\ 
\hline
Jj channel length & 300 nm\\ 
Jj channel width & 1 $\mu$m\\ 
Electron density & 2.0x10$^{12}$ cm$^{-2}$\\
Electronic mobility & $2\times10^4$ cm$^2$/Vs\\
normal resistance & 59 $\Omega$\\
Mean free path & 340 nm\\
Disorder temperature & 2.8 K\\
Bloch-Gr\"{u}neisen temp. & 76 K\\
$I_c(T_0)R_n$ product & 142 $\mu$eV\\
Thouless energy & 1.2 meV\\
Jj coupling energy & 2.0 meV\\
Effective capacitance & 3.65 fF \\
Plasma Freq. at zero bias current & $\leq$ 904 GHz\\
McCumber parameters & 0.23\\
NbN superconducting gap & 1.52 meV\\
\hline
\end{tabular}
\label{JJ}
\caption{List of Josephson junction parameters of the bolometer in this report.}
\end{table}

\bibliographystyle{nature}

\end{document}